\documentclass[aps,prl,article,groupedaddress,showpacs,twocolumn]{revtex4}
\usepackage{makeidx}
\usepackage{amssymb}
\usepackage{latexsym}
\usepackage{graphicx}
\makeindex
\begin{document}
\def\beq{\begin{equation}}
\def\beql#1{\begin{equation}\label{#1}}
\def\eeq{\end{equation}}
\def\bay{\begin{eqnarray}}
\def\bayl#1{\begin{eqnarray}\label{#1}}
\def\eay{\end{eqnarray}}
\def\r{\rho}\def\ro{\hat\r}\def\roe{\ro^{e}}
\def\qo{\hat q}\def\qoe{\qo^{e}}
\def\E{\mbox{E}}
\def\s{\sigma}\def\so{\hat\s}
\def\Ho{\hat H}
\def\g{\gamma}
\def\ket#1{\vert#1\rangle}
\def\bra#1{\langle#1\vert}
\newcommand\qexp[2]{\langle#1\rangle_{#2}}
\def\half{{\scriptstyle{\frac{1}{2}}}}
\def\tr{\mbox{tr}}
\def\roc{\ro^c}\def\roce{\ro^{ce}}
\def\qoc{\qo^c}

\title{Coupled Ito equations of continuous quantum state measurement, 
and estimation}
\author{Lajos Di\'osi$^a$, Thomas Konrad$^{b,c}$, 
  Artur Scherer$^b$ and J\"urgen Audretsch$^b$\\
$^a$Research Institute for Particle and Nuclear Physics, 
H-1525 Budapest 114, P.O.Box 49, Hungary
\\
$^b$Fachbereich Physik, Universit\"at Konstanz, Fach M 674,
D-78457 Konstanz, Germany\\
$^c$School of Pure and Applied Physics, University of KwaZulu-Natal, 
Durban 4000, South Africa}
\begin{abstract}
We discuss a non-linear stochastic master equation that governs the time-evolution
of the estimated quantum state. Its differential evolution corresponds to the infinitesimal
updates that depend on the time-continuous measurement of the true quantum state. The new
stochastic master equation couples to the two standard stochastic differential equations
of time-continuous quantum measurement. For the first time, we can prove that the calculated
estimate almost always converges to the true state, also at low-efficiency measurements. We show
that our single-state theory can be adapted to weak continuous ensemble measurements as well. 
\end{abstract}

\maketitle

{\it Introduction.\/}
For seven decades after the completion of quantum theory, sequential measurements on a single
quantum system were not technically accessible. Advancements of 
experimental technology have finally allowed multiple measurements under full control.
Time-continuous measurements on a single system have also become feasible. Their 
theory concluded to a flexible Ito-stochastic calculus. 
It describes the time evolution of the measured state $\ro_t$ under continuous measurement of a certain
variable $\qo$ as well as the evolution of the time-dependent measurement signal $q_t$. 
Recent perspectives of feed-back control shed new light on the problem. The time-continuous, 
or real-time, state determination became an immediate theoretical task. 
Doherty et al.~\cite{Doh99} worked out a theory for a specific case (cf.\ an application
for feedback induced cooling~\cite{Ste06}). A related different task has been discussed independently
\cite{SilJesDeu05}.
Our present proposal, sketched already in Ref.~\cite{Dio02}, extends
the results of \cite{Doh99} for 
the whole class of systems under (time-)continuous measurement, and we prove the general convergence of 
the real-time estimate to the true state. Our concept will be slightly different from that of Ref.~\cite{Doh99}. 
We do not think of integrating the stochastic differential equation of continuous measurement. Rather, we emphasize
that the estimated state $\ro_t^e$ satisfies a further Ito-stochastic equation driven
by the (noisy) measurement signal $q_t$. This allows us to prove that the Hilbert-Schmidt distance between
the unknown state $\ro_t$ and the calculated real-time estimate $\ro_t^e$ is
decreasing until $\ro_t$ and $\ro_t^e$ will coincide:
\beql{conv}
\lim_{t\rightarrow\infty}\Vert\ro_t^e-\ro_t\Vert=0.
\eeq
This convergence implies for example the possibility to monitor
quantum oscillations in real-time (cf.~\cite{AudKleeKon04}).
The structure of our Letter is the following. We first introduce the elementary measurement-update cycle and the weak-measurement
limit. Then we present the Ito equations of our proposal, and we provide the proof of the above convergence theorem.
We discuss and verify the theory also for non-maximum efficiency of
the continuous measurement. Finally, an application to
collective measurements is shown. 

{\it Unsharp measurement and update of states.\/}
In all continuous measurement and/or estimation theories, unsharp
measurements play a central role. 
Consider the standard Gaussian model of unsharp measurement of a variable $\qo$ 
\cite{CavMil87}. If $\ro$ is the a priori state the measurement outcome $q$ will
have the following probability distribution:
\beql{unsh1}
p(q)=\tr\left[ G_\s(q-\qo)\ro \right]=\qexp{G_\s(q-\qo)}{\ro},
\eeq
where $G_\s$ is the normalized Gauss function of spread $\s$.
The following standard update yields our a posteriori state:
\beql{unsh2}
\ro\longrightarrow\frac{1}{p(q)}G_\s^{1/2}(q-\qo)\ro G_\s^{1/2}(q-\qo).
\eeq
If the a priori state $\ro$ is unknown then also the a posteriori one remains
unknown. We can, nonetheless, estimate the a priori state, say by a certain $\roe$.
Then we apply the {\it same} update to our a priori estimate $\roe$ as to the true state above:
\beql{unsh3}
\roe\longrightarrow\frac{1}{p^e(q)}G_\s^{1/2}(q-\qo)\roe G_\s^{1/2}(q-\qo).
\eeq
Please note that the normalization factor is different than in
Eq.~(\ref{unsh2}). 
The normalizing function $p^e(q)$ 
has, although  we employ similar notation, no role as a probability distribution. 
We expect that by using weak measurements, i.e. when the unsharpness $\s$ is very
large, iterating the updates (\ref{unsh1}-\ref{unsh3}) brings the estimate and the true 
state closer to each other. It therefore makes sense to repeat the above measurement-update-cycle 
many times at high frequency $\nu$ in order to
make the real-time estimate $\roe_t$ converge to the true state $\ro_t$, as
claimed in Eq.(\ref{conv}).
Below we prove this remarkable convergence in the weak-measurement continuous-time limit \cite{Dio06}, 
where both the unsharpness
$\s$ and the repetition frequency $\nu$ of the measurement-update cycle tend to infinity
while the ratio $\nu/\s^2$ remains constant:
\beql{weak}
\s,~\nu~~~\longrightarrow~~\infty,~~~~~~~\frac{\nu}{\s^2}=\g.
\eeq
The quantity $\g$ is called the strength of the continuous measurement. In this limit, 
Eqs.~(\ref{unsh1}-\ref{unsh3}) result in three Ito stochastic differential equations, respectively,
for the time-dependent outcome (signal) $q_t$ of the measurement, for the true state $\ro_t$ and for
the estimate $\roe_t$, which constitute the theory of continuous measurement and estimation. 

{\it The three coupled Ito equations.\/}
Let us first postulate the heuristic theory.
Consider an observable $\qo$ which we measure continuously. 
The signal is governed by  a simple stochastic equation:
$
q_t=\qexp{\qo_t}{\ro_t}+\g^{-1/2}w_t
$
where $\qexp{\qo_t}{\ro_t}$ stands for the mean value of $\qo$ in the current quantum state $\ro_t$. The $w_t$
is the standard white-noise defined by
$\E[w_t]=0$ and $\E[w_t w_s]=\delta(t-s)$ where $\E$ stands for the stochastic mean.
This form of the observed value $q_t$ is plausible: it fluctuates around the quantum mean value
and the magnitude of the fluctuations is proportional to the strength of the continuous measurement. 
Due to its non-linearity, however, the naive equation must be replaced by the mathematically precise 
Ito equation:
\beql{Ito1}
dQ=\qexp{\qo}{\ro} dt+\g^{-1/2}dW
\eeq
where $Q_t,W_t$ are the time-integrals of $q_t$ and $w_t$, respectively. From now on, we call $Q_t$ the
integrated signal of the continuous measurement. For notational convenience of Eq.(\ref{Ito1}) 
and of further equations, the lower indices $t$ are systematically ignored.  

The second Ito equation governs the state $\ro_t$ under continuous measurement of $\qo$. 
For Markovian mechanisms, like ours, the Ito increment $d\ro_t$ of the state will only
depend on the current state $\ro_t$ and on the current 
Ito increment  $dQ_t$ of the (integrated) signal \cite{foo1}:
\begin{eqnarray}\label{Ito2}
d\ro=&&-i[\Ho,\ro]dt-\frac{\g}{8}[\qo,[\qo,\ro]]dt\nonumber\\
     &&+\frac{\g}{2}\{\qo-\qexp{\qo}{\ro},\ro\}\left(dQ-\qexp{\qo}{\ro}dt\right),
\end{eqnarray}
where $\Ho$ is the Hamiltonian.
Obviously, at any time $t$, $\ro_t$ is  conditioned on the previous measurement outcomes through
$\{Q_s;s\leq t\}$.

The third Ito equation governs the evolution of our estimate $\roe_t$:
\begin{eqnarray}\label{Ito3}
d\roe=&&-i[\Ho,\roe]dt-\frac{\g}{8}[\qo,[\qo,\roe]]dt\nonumber\\
      &&+\frac{\g}{2}\{\qo-\qexp{\qo}{\roe},\roe\}\left(dQ-\qexp{\qo}{\roe}dt\right)\,. 
\end{eqnarray}
The structure of this equation coincides with the structure of Eq.~(\ref{Ito2}), but $dQ$ in here depends
on the true state $\ro$, cf.~Eq.~(\ref{Ito1}). 

The Eqs.~(\ref{Ito1}-\ref{Ito3}) constitute the theory of continuous measurement {\it and} estimation. 
The second and third can also be called the stochastic master equation (SME) of measurement and estimation, respectively.
The first two constitute the theory of continuous measurement and they were shortly derived from  
unsharp measurements (\ref{unsh1}-\ref{unsh3}) in the weak-measurement
 continuous-time limit, cf. Ref.~\cite{Dio88}. 
The proof relies on the approximation \beql{unsh1appr}
p(q)\approx G_\s(q-\qexp{\qo}{\ro})
\eeq
valid for large $\s$. We can easily confirm the novel SME (\ref{Ito3}) from Eq.~(\ref{unsh3}), 
without adapting the (otherwise simple) derivation \cite{Dio88} 
of (\ref{Ito1},\ref{Ito2}) from Eqs.~(\ref{unsh1},\ref{unsh2}). 
By change of variables, we are going to show that (\ref{unsh2}) and (\ref{unsh3}) become asymptotically identical. 
Let us consider the expression (\ref{unsh3}) of the updated estimate and calculate 
the normalizing denominator for large $\s$:
\beql{unsh3appr1}
p^e(q)\approx G_\s(q-\qexp{\qo}{\roe}).
\eeq
Let us introduce new variables: $\qoe=\qo-\qexp{\qo}{\roe}+\qexp{\qo}{\ro}$ and, of course,
$q^e=q-\qexp{\qo}{\roe}+\qexp{\qo}{\ro}$. Please observe that $p^e(q)=p(q^e)$ and re-write (\ref{unsh3}) into
this form:
\beql{unsh3appr2}
\roe\rightarrow\frac{1}{p(q^e)}G_\s^{1/2}(q^e-\qoe)\roe G_\s^{1/2}(q^e-\qoe).
\eeq
This is exactly the same equation as equation (\ref{unsh2}) updating the true state $\ro$. 
Therefore the SME for $\roe$ will, in the new variables $\qoe,q^e$, coincide with the 
SME (\ref{Ito2}) of $\ro$:
\begin{eqnarray}\label{Ito2e}
d\roe=&&-i[\Ho,\roe]dt-\frac{\g}{8}[\qoe,[\qoe,\roe]]dt\nonumber\\
     &&+\frac{\g}{2}\{\qoe-\qexp{\qoe}{\roe},\roe\}\left(dQ^e-\qexp{\qoe}{\roe}dt\right).
\end{eqnarray}
If we restore the original variables $\qo=\qoe+\qexp{\qo}{\roe}-\qexp{\qo}{\ro}$
and $dQ=dQ^e+\qexp{\qo}{\roe}dt-\qexp{\qo}{\ro}dt$, we obtain Eq.~(\ref{Ito3}).

{\it Proof of convergence.\/}
For long times, both the actual state of the system, $\ro_t$,  and the estimated state $\roe_t$ become pure states.
Therefore it will be sufficient to prove that the fidelity $\tr[\ro_t\roe_t]$ tends to $1$ for
large $t$. We detail the proofs below.

First we note that all three equations (\ref{Ito1}-\ref{Ito3}) 
are invariant under the trivial shifts
$\qo\rightarrow\qo+r$, $dQ\rightarrow dQ+rdt$, for arbitrary
real constant $r$. 
In all time-local calculations we can, e.g., make 
$\qexp{\qo}{\ro}$ zero by choosing $r=-\qexp{\qo}{\ro}$ and we can restore the true result at the end of 
the calculation if we make a second shift by $-r$. This allows quicker calculations and we shall refer to 
this as shift invariance. 

For long times the solutions $\ro_t$ of the SME (\ref{Ito1}) are known to turn into pure states
\cite{Doh99,Kor00}. To prove this, we show that the increment of $\E\left[\tr[\ro_t^2]\right]$ is non-negative:
\beql{droro}
d\E\Big[\tr[\ro^2]\Big]=\E\Big[\tr[2\ro d\ro + d\ro d\ro]\Big]\geq0.
\eeq
Using the Ito equations (\ref{Ito1},\ref{Ito2}) in the shifted coordinate system where
$\qexp{\qo}{\ro}=0$, the l.h.s.\ can be written as the trace of a nonnegative matrix:
\beql{droro1}
\g^{-1}\frac{d}{dt}\E\Big[\tr[\ro^2]\Big]=\tr\left[\ro\qo\ro\qo\right]\equiv\tr\Big[(\ro^{1/2}\qo\ro^{1/2})^2\Big] \,,
\eeq
which is greater than zero and vanishes if $\ro_t$ is already a pure state. 
In certain marginal cases the growth of purity may get stalled, we
discuss this problem later.  However, in 
generic physical situations the r.h.s.\ of Eq.~(\ref{droro1}) 
becomes zero only if $\ro_t$ turns into a pure state. 
The presented proof implies the longtime purity of $\roe_t$ as well since, in suitable variables, 
the SMEs of $\ro_t$ and $\roe_t$ are identical, cf. Eqs.(\ref{Ito2}) and (\ref{Ito2e}), respectively.

Finally we prove that, in the generic case,  
$\tr[\ro_t\roe_t]\rightarrow1$ when $t\rightarrow\infty$.  
The task is to show that
\beql{droero}
d\E\Big[\tr[\ro\roe]\Big]=\E\Big[\tr[d\ro \roe + \ro d\roe + d\ro d\roe]\Big]\geq0\,,
\eeq
with equality if and only if $\roe=\ro$ for all typical, 
physically interesting situations.
Using shift invariance, we can set $\qexp{\qo}{\ro}+\qexp{\qo}{\roe}=0$. 
Substituting Eqs.~(\ref{Ito2},\ref{Ito3}) yields  the following result:
\begin{eqnarray}\label{droero1}
&&\g^{-1}\frac{d}{dt}\E\Big[\tr[\ro\roe]\Big]=\\
&=&\qexp{\qo}{\ro}^2\,\tr\left[\ro\roe\right]+\tr\left[\ro\qo\roe\qo\right]
  +\qexp{\qo}{\ro}\,\tr\left[\qo\{\ro,\roe\}\right]\nonumber\\
&\equiv&\tr\left[\ro^{1/2}(\qo+\qexp{\qo}{\ro})\roe(\qo+\qexp{\qo}{\ro})\ro^{1/2}\right]\nonumber \,.
\end{eqnarray}
The r.h.s.\ is the trace of a nonnegative matrix. This assures that the fidelity is monotonously increasing until 
$\roe=\ro$ is reached asymptotically. We shall emphasize that the
  convergence may cease for marginal cases, 
see our discussion below. Nevertheless, in generic physical
applications convergence will always be achieved. 
This claim is supported by numerical simulations, and also 
by the fact that in the one-qubit case, if $[\Ho,\qo]\neq0$,
one can exactly prove that the  r.h.s.\ of Eq.~(\ref{droero1})
can vanish only if the true and the estimated state coincide, 
$\roe=\ro$.

{\it Extension for low-efficiency measurements.\/} 
The theory of continuous measurement (\ref{Ito1},\ref{Ito2}) corresponds to perfectly efficient continuous
measurements, i.e.,\ the signal-to-noise ratio reaches the quantum mechanically possible maximum value.
Real continuous measurements are producing and/or are accompanied by an excess noise. 
Therefore they cannot preserve or reach the purity of continuously measured states although they limit
their mixedness. Wiseman and Milburn \cite{WisMil93} have already incorporated the efficiency parameter $\eta\in[0,1]$ 
into the theory (\ref{Ito1},\ref{Ito2}), and we do it for the novel SME (\ref{Ito3}) as well:
\begin{eqnarray}
dQ&=&\qexp{\qo}{\ro} dt+(\eta\g)^{-1/2}dW\label{Ito1loss}\\
d\ro&=&-i[\Ho,\ro]dt-\frac{\g}{8}[\qo,[\qo,\ro]]dt\nonumber\\
     &&+\frac{\eta\g}{2}\{\qo-\qexp{\qo}{\ro},\ro\}\left(dQ-\qexp{\qo}{\ro}dt\right),\label{Ito2loss}\\
d\roe&=&-i[\Ho,\roe]dt-\frac{\g}{8}[\qo,[\qo,\roe]]dt\nonumber\\
      &&+\frac{\eta\g}{2}\{\qo-\qexp{\qo}{\roe},\roe\}\left(dQ-\qexp{\qo}{\roe}dt\right)\,.\label{Ito3loss} 
\end{eqnarray}
Our SME (\ref{Ito3loss}) of estimation works for lower efficiencies
$\eta<1$ as well. We expect that the convergence
(\ref{conv}) of the estimate $\roe_t$  and the true state $\ro_t$ will slow down if $\eta\ll1$, 
still it exists for all nonzero efficiencies $\eta$. The former proof cannot be applied directly because it 
relied upon the longtime purity of both $\ro_t$ and $\roe_t$. Yet, we can reduce the proof of the case
$\eta<1$ to the former proof of the case $\eta=1$. Let us outline the steps.

In the two SMEs (\ref{Ito2loss},\ref{Ito3loss}), 
let us separate the excess noise from that which is necessary for a given measurement efficiency $\eta$:
\begin{eqnarray}\label{Ito2lossexcess}
d\ro=&&-i[\Ho,\ro]dt-\frac{(1-\eta)\g}{8}[\qo,[\qo,\ro]]dt-\frac{\eta\g}{8}[\qo,[\qo,\ro]]dt\nonumber\\
     &&+\frac{\eta\g}{2}\{\qo-\qexp{\qo}{\ro},\ro\}\left(dQ-\qexp{\qo}{\ro}dt\right)\,,
\end{eqnarray}
and similarly for (\ref{Ito3loss}).
It is known that the noise term proportional to $1-\eta$, like any further noise terms, can always be reproduced 
formally by an interaction Hamiltonian with a ``heat bath''. Accordingly, we can transform the original SMEs
of continuous measurement and estimation at measurement strength $\g$ and efficiency $\eta<1$ into the theory of
continuous measurement and estimation of the system+bath at measurement strength $\eta\g$ and efficiency $\eta=1$. 
Vice versa, if we trace over the bath, these SMEs reduce to the SMEs of the original system. 
According to our earlier theorem, valid for $\eta=1$, the true and the estimated state of the 
system+bath converge to each other. This convergence implies, via tracing over the bath, 
that also $\ro_t$ and $\roe_t$ converge to each other whereas both may remain mixed forever.

{\it Application to ensembles.\/} 
The weak measurement paradigm also plays a role in applications where
a large ensemble of the unknown state    
$\ro$ is accessible to the experiment,
cf. e.g. Ref.\cite{SilJesDeu05}. Crucial in this context is 
the approximation that the collective state of the ensemble preserves the uncorrelated form  
$\roc_t=\ro_t^{\otimes N}$ if $N$ is very large and the strength $\g^c$ of the collective measurement 
is very small. 
In particular, we consider the same observable $\qo$ on each component and we measure their 
{\it sum} $\qoc$ in a collective continuous measurement of strength $\g^c$.   
For simplicity's sake only, we assume $\eta=1$ and apply the theory (\ref{Ito1}-\ref{Ito3}):
\begin{eqnarray}
dQ^c&=&\qexp{\qo^c}{\roc} dt+(\g^c)^{-1/2}dW\,,\label{Ito1ens}\\
d\roc&=&-i[\Ho^c,\roc]dt-\frac{\g^c}{8}[\qoc,[\qoc,\roc]]dt\nonumber\\
     &&+\frac{\g^c}{2}\{\qoc-\qexp{\qoc}{\roc},\roc\}\left(dQ^c-\qexp{\qoc}{\roc}dt\right)\,,\label{Ito2ens}\\
d\roce&=&-i[\Ho^c,\roce]dt-\frac{\g^c}{8}[\qoc,[\qoc,\roce]]dt\nonumber\\
      &&+\frac{\g^c}{2}\{\qoc-\qexp{\qoc}{\roce},\roce\}\left(dQ^c-\qexp{\qoc}{\roce}dt\right) \;\label{Ito3ens} 
\end{eqnarray}
where $\Ho^c$ is the collective Hamiltonian, i.e., the sum of the same $\Ho$ for all $N$ components. 
We extend the approximation $\roc=\ro^{\otimes N}$ for 
the estimate: $\roce=(\roe)^{\otimes N}$. Substituting these forms, we obtain closed equations of
the ensemble-continuous-measurement and single-system-estimation: 
\begin{eqnarray}
dQ^c&=&N\qexp{\qo}{\ro} dt+(\g^c)^{-1/2}dW\,,\label{Ito1ensred}\\
d\ro&=&-i[\Ho,\ro]dt-\frac{\g^c}{8}[\qo,[\qo,\ro]]dt\nonumber\\
     &&+\frac{\g^c}{2}\{\qo-\qexp{\qo}{\ro},\ro\}\left(dQ^c-N\qexp{\qo}{\ro}dt\right)\,,\label{Ito2ensred}\\
d\roe&=&-i[\Ho,\roe]dt-\frac{\g^c}{8}[\qo,[\qo,\roe]]dt\nonumber\\
      &&+\frac{\g^c}{2}\{\qo-\qexp{\qo}{\roe},\roe\}\left(dQ^c-N\qexp{\qo}{\roe}dt\right)\,.\label{Ito3ensred} 
\end{eqnarray}
These equations are identical to the Eqs.(\ref{Ito1}-\ref{Ito3}) of continuous measurement and
estimation on a single system, apart from two things. First, the strength $\g^c$ of the collective 
measurement survives as the strength of the single state measurement. Second, the SMEs are governed
by the collective signal $Q_t^c$, as they should be. This latter fact leads usually to a faster convergence
of $\ro_t$ and $\roe_t$ than the single state method, as it is plausible and could be verified from a 
detailed analysis.  

{\it Remarks.\/}
As we anticipated in the text, there are exceptions from convergence (\ref{conv}) and from
longtime purity of $\ro_t$ and $\roe_t$.
For trivial dynamics $[\Ho,\qo]=0$, the estimate $\roe_t$ will get stuck in any eigenstate of $\qo$, 
independently of $\ro_t$ which would converge to any other eigenstate as time goes by. 
However, these cases are of marginal importance. In real tasks the
dynamics is nontrivial and $[\Ho,\qo]$ does not vanish. We conjecture the following condition as sufficient
for the universal convergence. Consider the {\it Heisenberg} operator $\qo_t$ of the measured observable,
and determine the largest common eigenspace of all $\qo_t$ for $t\geq0$. If this eigenspace is empty
or one-dimensional then the convergence of $\ro_t$ and $\roe_t$ is always guaranteed.  
For instance, the position measurement of a particle yields a convergent state estimate in one
dimension. The two-dimensional motion may be different. For a free particle, the simultaneous 
continuous measurement of both coordinates $\qo_x$ and $\qo_y$ is necessary otherwise the measured state 
may not become pure and the estimate may not converge to it. Interestingly, there is a chance of
purity and convergence if we measure but one coordinate $\qo_x$, provided a potential rotates $\qo_x$'s
Heisenberg-version in a proper non-trivial way. 

In practice, the (integrated) signal $Q_t$ is obtained from the experimental device doing the
continuous measurement, so that $Q_t$ does not need computational efforts. On the other hand,
the real-time estimate $\roe_t$ must be on-line calculated from $Q_t$ and one is interested in
good algorithms. There are several options, depending on the concrete task. In case of
optimum detection efficiency $\eta=1$, we can use a pure state estimate 
from the beginning. Then the density matrix equation (\ref{Ito3}) 
is equivalent to the following stochastic Schr\"odinger equation for the state vector estimate:  
\begin{eqnarray}\label{Ito3Sch}
d\psi^e=&&-i\Ho\psi^e dt-\frac{\g}{8}\left(\qo-\qexp{\qo}{\psi^e}\right)^2\psi^e dt\nonumber\\
       &&+\frac{\g}{2}\left(\qo-\qexp{\qo}{\psi^e}\right)\psi^e\left(dQ-\qexp{\qo}{\psi^e}dt\right)\,. 
\end{eqnarray}
Of course, if we calculate $d\roe_t=d[\psi^e_t(\psi^e_t)^\dagger]$ from the above stochastic 
Schr\"odinger equation, we get back (\ref{Ito3}).

{\it Summary.\/}
To complete the standard theory of continuous measurement, we have constructed a third Ito
stochastic equation for the real-time state estimate, exploiting the measured signal. Our theory
(\ref{Ito1loss}-\ref{Ito3loss}) 
applies to any system under time-continuous measurement. In this way, we have largely extended 
similar heuristic proposals \cite{Doh99,Ste06} which used Gaussian estimates $\roe_t$ and requested,
in principle, the perfect efficiency (signal-to-noise) of the time-continuous measurement.
We proved analytically that our novel SME for the state estimate yields the true state for any non-trivial 
dynamics, at any nonzero efficiency of the measurement. The recent work \cite{Dio02} on real-time estimate 
has sketched the Ito equations in an alternative representation, without details of derivation
and proof of the estimate's convergence to the true state. Our theory, due to the plain structure of 
the equations, can invariably be applied when several observables are measured simultaneously, 
like the canonical coordinate and momentum $\qo,\hat p$ or the spatial coordinates $\qo_x,\qo_y,\qo_z$
of a particle, as well as two or more components of a Pauli spin. We also showed that the theory 
applies when the state estimate relies on the collective continuous measurement on a large number
of copies. Our SME applies to the experimental realizations of single state control, and 
we expect that it will contribute to a direct solution of state tomography from continuous measurement 
on ensembles, cf. e.g. Ref.\cite{SilJesDeu05}.

L.D. thanks the Hungarian OTKA Grant 49384 and the Center for Applied
Photonics (CAP) at the University of Konstanz for its hospitality. This work
was supported by CAP.


\begin{thebibliography}{99}
\bibitem{foo1} The evolution equation for the measured state $\ro_t$ can be equivalently
written as an autonomous equation \cite{Gis84,Dio88} driven by the standard Wiener-process $W_t$: 
$
d\ro=-i[\Ho,\ro]-\frac{1}{8}g^2[\qo,[\qo,\ro]]dt+\frac{1}{2}g\{\qo-\qexp{\qo}{\ro},\ro\}dW.
$
This form will be convenient for analytic calculations.
\end{thebibliography}
\end{document}